# A Finger on the Pulse of Cardiovascular Health: Estimating Blood Pressure with Smartphone Photoplethysmography-Based Pulse Waveform Analysis

Ivan Liu[1,2], Fangyuan Liu[1], Qi Zhong[2], Shiguang Ni[3]

1 Department of Psychology, Faculty of Arts and Sciences, Beijing Normal University at Zhuhai, Zhuhai, Guangdong, China
2 Faculty of Psychology, Beijing Normal University, Beijing, China
3 Shenzhen International Graduate School, Tsinghua University, Shenzhen, China

Corresponding Author: Shiguang Ni[3]
University Town of Shenzhen, Nanshan District, Shenzhen, 518055, China
Email address: ni.shiguang@sz.tsinghua.edu.cn

**Abstract:**

Utilizing mobile phone cameras for continuous blood pressure (BP) monitoring presents a cost-effective and accessible approach, yet it is challenged by limitations in accuracy and interpretability. This study introduces four innovative strategies to enhance smartphone-based photoplethysmography for BP estimation (SPW-BP), addressing the interpretability-accuracy dilemma. First, we employ often-neglected data-quality improvement techniques, such as height normalization, corrupt data removal, and boundary signal reconstruction. Second, we conduct a comprehensive analysis of thirty waveform indicators across three categories to identify the most predictive features. Third, we use SHapley Additive exPlanations (SHAP) analysis to ensure the transparency and explainability of machine learning outcomes. Fourth, we utilize Bland-Altman analysis alongside AAMI and BHS standards for comparative evaluation. Data from 127 participants demonstrated a significant correlation between smartphone-captured waveform features and those from standard BP monitoring devices. Employing multiple linear regression within a cross-validation framework, waveform variables predicted systolic blood pressure (SBP) with a mean absolute error (MAE) of 9.86±6.78 mmHg and diastolic blood pressure (DBP) with an MAE of 8.01±5.15 mmHg. Further application of Random Forest models significantly improved the prediction MAE for SBP to 8.91±6.30 mmHg and for DBP to 6.68±4.54 mmHg, indicating enhanced predictive accuracy. Correlation and SHAP analysis identified key features for improving BP estimation. However, Bland-Altman analysis revealed systematic biases, and MAE analysis showed that the results did not meet AAMI and BHS accuracy standards. Our findings highlight the potential of SPW-BP, yet suggest that smartphone PPG technology is not yet a viable alternative to traditional medical devices for BP measurement.

## 1.Introduction

The circulatory system's importance in delivering oxygen and nutrients to tissues and removing waste underscores the necessity of maintaining balanced blood pressure (BP) for health. High BP is associated with serious conditions such as heart attacks, strokes, kidney complications, and eye diseases (Katsanos et al., 2017). Routine BP monitoring is essential; however, readings vary due to factors such as time of day, exercise regimen, diet, mental state, social triggers, measurement posture, and environment (Schutte et al., 2022). BP values can also show substantial variability between individual heartbeats (Forouzanfar et al., 2015), necessitating multiple daily measurements.

Continuous BP monitoring, however, faces significant challenges, such as limited access to medical-grade equipment (Athaya & Choi, 2022), the need to carry portable cuff BP devices all day, and additional costs. Furthermore, sphygmomanometer-based BP devices can cause discomfort and disrupt blood flow during measurement (O'Brien, 2003). Additionally, the COVID-19 pandemic has further complicated access to medical devices. To overcome these challenges, the use of smartphones for daily BP monitoring is being explored (Park et al., 2020; Sharma et al., 2022).

Smartphones, through their built-in sensors, can measure BP by two principal methods (El-Hajj & Kyriacou, 2020). The Pulse Arrival Time (PAT) method estimates the speed of the heartbeat pressure wave but necessitates an ECG for pulse onset detection (Poon & Zhang, 2006), potentially requiring additional devices (Nemcova et al., 2020). Despite exploration into PAT estimation with a single smartphone, this method lacks wide acceptance (Tabei et al., 2020; Yoon et al., 2023). Pulse Waveform Analysis (PWA), on the other hand, determines BP based on the pulse waveform shape. Arterial stiffness causes reflected waves within the vessel wall to arrive more rapidly (Figure 1), leading to a pronounced increase in BP (Baruch et al., 2011).

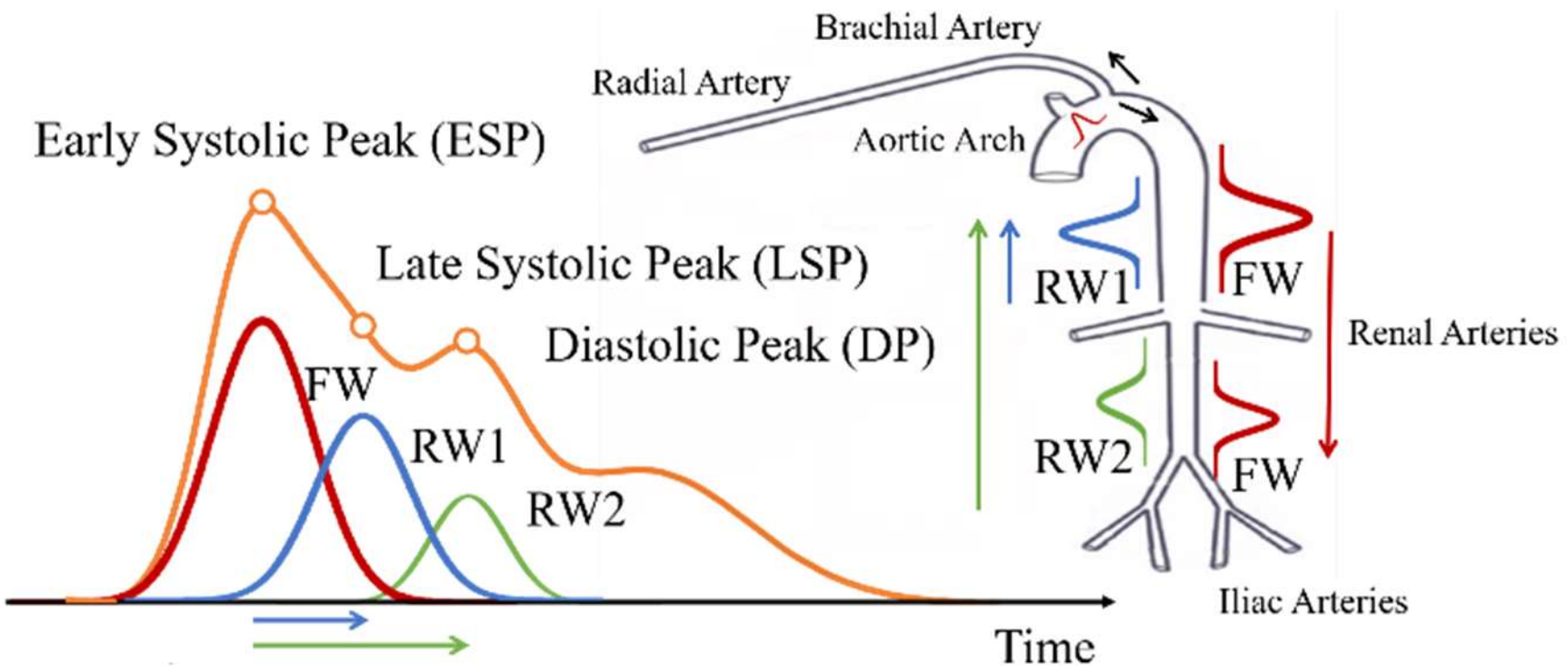

Figure 1. The reflected waves and the shape of the radial arterial pulse waveform (adapted from (Baruch et al., 2011))

Photoplethysmography (PPG) technology, which leverages optical sensors, is widely used to detect waveform changes by monitoring subcutaneous blood volume variations. By analyzing the varying intensities of reflected light, differences in hemoglobin absorption can be calculated, enabling the assessment of blood volume increases (systolic phase) or decreases (diastolic phase). Traditional LED-based PPG analysis commonly employs finger oximeters, whereas smartphone-based PPG measures heartbeats via the phone's camera (Jonathan & Leahy, 2011). Numerous studies have documented the effectiveness of LED-based PPG methods in BP estimation (Radha et al., 2019), and several have combined PAT and PPG to enhance BP prediction accuracy (Sun et al., 2016).

Despite its promise, smartphone PPG-based waveform analysis for blood pressure prediction (SPW-BP) has yet to receive adequate attention in research and practical applications. A recent systematic review by Frey et al. (2022) identified only twenty-five studies on smartphone PPG-based BP monitoring, with just fifteen employing pulse wave analysis methods.

Current PWA-based BP measurements have not garnered the necessary attention for several reasons. Firstly, the measurement accuracy of BP estimation from smartphone PPG has raised concerns among scholars. Besides the inferior signal quality of smartphone PPG, researchers are concerned that the complex nature of peripheral vascular resistance in the carotid artery may render smartphone PPG-based BP estimates unsuitable for practical applications, as studies have shown (Bruyndonckx et al., 2013). Secondly, to address the issue of low signal quality, pioneering SPW-BP researchers have utilized machine learning to enhance prediction accuracy (Haugg et al., 2022). Although studies using machine learning for BP assessment have surpassed traditional methods in the measurement accuray, the limited interpretability of these models hinders their acceptance outside academia (Murdoch et al., 2019), particularly in medical decision-making where transparency and patient understanding are paramount (London, 2019; Vellido, 2019).

Thirdly, studies have employed various waveform features to estimate BP, but many simply report using features established by predecessors without a comprehensive rationale for their selection. This lack of consistency and absence of thorough comparative analysis and feature selection process significantly obstructs the integration of past research into a unified knowledge base, thereby impeding progress in the field. Lastly, there has been a notable oversight regarding the need for proper benchmarking against reliable, gold-standard metrics, including the AAMI and BHS standards and Bland-Altman analysis.

Our study aims to bridge the gap in existing research by addressing the accuracy-interpretability dilemma in BP measurement using smartphone PPG methodology. Firstly, our study emphasizes the importance of robust data preprocessing to improve data quality, challenging the trend observed in many SPW-BP studies of favoring complex models over basic data preparation (Athaya & Choi, 2022). These techniques—normalization with body height, removal of low-quality samples, boundary data reconstruction, and collinearity elimination—collectively improve data integrity and the effectiveness of subsequent analyses.

Secondly, our research conducts an extensive examination of cardiac waveform features, as documented in existing literature, to elucidate their correlations with BP and evaluate their predictive value. Additionally, we explore the incorporation of interpretable machine learning methodologies in waveform-based BP analysis to investigate feature importance. Despite black-box machine learning models generally offering superior predictive capabilities, they frequently lack the interpretability necessary for medical applications (Molnar et al., 2020). To improve the understandability of these models and to identify key features for further analysis, we utilize SHapley Additive exPlanations (SHAP) for visualizing crucial features in BP prediction (Lundberg & Lee, 2017).

Lastly, our study applies the Bland-Altman method, a benchmark in medical engineering, to assess our BP prediction technique's congruence with standard ECG references. Following Qin et al., (2023), the results are reported using the AAMI and BHS standards to make the findings comparable with other medical equipment.

# 2.Methods

## 2.1.Data collection

Data were collected from 127 students and university employees in Shenzhen, China, comprising 56.69% males, with an average age of 22.78 years (SD = 1.97). Of these, 113 provided valid pulse waveform data, with 59.29% being male and an average age of 22.74 years (SD = 1.83). The study did not specify any inclusion or exclusion criteria. The study protocol was approved by the ethical board of the department of psychology at Tsinghua University, and informed consent was obtained from all participants. Participants were instructed to hold a smartphone (Mi-8 SE, Xiao-Mi, China) in their left hand and use the Heartily Happy app (HH, developed by our team and available on the Google Play Store), to record six 4-minute videos of their fingertips at a resolution of 120×160 pixels. Before and after the data collection session, participants measured their BP using a cuff pressure monitor (HEM-6322T, Omron, Japan); the average of these two measurements served as the reference for comparison.

## 2.2.Signal Processing

### 2.2.1.Convert RGB signals to waveform

In each data collection session, the HH app activates the device's built-in flashlight to record videos at a resolution of 120 × 160 pixels, capturing around 30 frames per second, as Liu et al. (2020) detailed (Figure 2). Raw picture frames, in YUV format retrieved from the preview function, are then converted into RGB format. The input signals $(r_i, b_i, g_i)$ from the red, blue, and green color channels undergo normalization, using a 100-point moving average $(\bar{R}, \bar{B}, \bar{G})$ and standard deviation $(\sigma_R, \sigma_B, \sigma_G)$ for processing:

$$n_R(r_i) = \frac{r_i - \bar{R}}{\sigma_R}, n_B(b_i) = \frac{b_i - \bar{B}}{\sigma_B}, n_B(g_i) = \frac{g_i - \bar{G}}{\sigma_G}, i \in \boldsymbol{A} = (1, \dots, k), \tag{1}$$

where $k$ is the number of frames in each data collection session. We then calculate the standard deviation-weighted average $f(t_i)$ as follows:

$$f(t_i) = \frac{\sigma'_G \times n_G(g_i) - \sigma'_B \times n_B(b_i) - \sigma'_R \times n_R(r_i)}{\sigma'_R + \sigma'_G + \sigma'_B}, \tag{2}$$

where

$$\sigma_C' = \begin{cases} \sigma_C & \text{if } \sigma_C > 0.5, 3 < \bar{C} < 252 \\ 0 & \text{otherwise} \end{cases}, C \in \{R, B, G\}$$

and $t_i \in \boldsymbol{T} = \{t_i | i \in 1, .., k\}$ denotes the time at which the i$^{\text{th}}$ data point was collected. A color channel is excluded from the $f(t_i)$ if the average of the input ($\bar{C}$) is either significantly low or approaches the upper limit of 255, or if its standard deviation ($\sigma_C$) is too small ($\sigma_C \leq 0.5$). Furthermore, the signs of the red and blue channels are inverted to reflect their inverse relationship with the green channel.

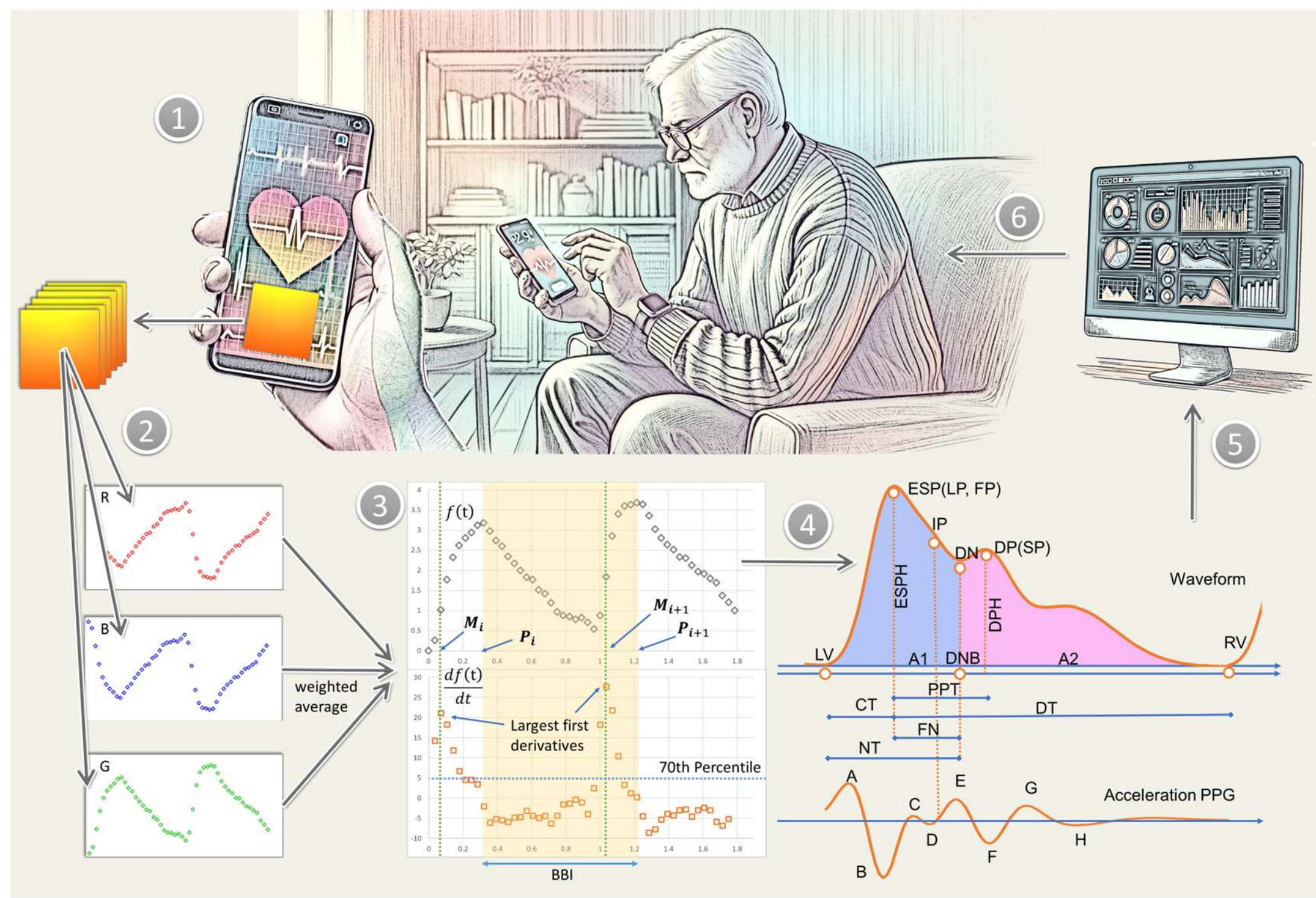

Figure 2 The data processing flow entails: (1) collecting data; (2) converting RGB signals to waveforms; (3) determining BBIs; (4) extracting waveform features; (5) using waveform features to predict BP; and (6) providing BP estimates to users [Adapted from Liu et al., (2020) and Elgendi et al., (2014)].

### 2.2.2.Beat-to-beat (BBI) segmentation

After converting signals into waveform inputs, $f(t_i)$ is segmented into intervals corresponding to each BBI. The study then identifies potential BBIs by using a set of local maxima ($\boldsymbol{M_i}$) – those surpassing the 70$^{\text{th}}$ percentile ($P_{70}$) of the first derivatives $f'(t_i)$:

$$\boldsymbol{M} = \{\boldsymbol{M_1}, \boldsymbol{M_2}, \dots, \boldsymbol{M_u}\} = \left\{ t_i | \max f'(t_i) > P_{70}, f''(t_i) < 0, \frac{60}{t_j - t_i} < 150 \;\; \forall j \neq i, t_i \in \boldsymbol{T} \right\}. \tag{3}$$

The distances between two successive data points in the set $\boldsymbol{M}$ are converted into HRs to exclude points exhibiting an HR exceeding 150 beats per minute (bpm). The intervals segmented by the data points in $\boldsymbol{M}$ are then defined as the set of BBIs ($\boldsymbol{B}$):

$$\boldsymbol{B} = \left\{(t_1, t_2) | t_1 \in \boldsymbol{M}, t_2 = \min\left\{t_i \in \boldsymbol{M} \,\middle|\, t_i > t_1, \frac{60}{t_i - t_1} < 150\right\}\right\}. \quad (4)$$

### 2.2.3.BBI normalization

To mitigate issues associated with baseline drift, the amplitude of data points within each BBI is normalized (Peng et al., 2015). This normalization is based on the height difference between two successive peaks, as outlined below:

$$f_{\text{normalized}}(t_i) = f(t_i) + \left(\frac{t_i - t_L}{t_R - t_L}\right) \times \left(f(t_R) - f(t_L)\right), \forall t_i \in \ (t_L, t_R), (t_L, t_R) \in \boldsymbol{B} \quad (5)$$

The current study then apply the R package RHRV (Martínez et al., 2017) to convert the heartbeat points generated by the fiducial point detection techniques to HRV measures for further analysis.

## 2.3.Waveform Metrics

This study utilizes three groups of waveform features commonly referenced in the literature (TABLE 1). Time and altitude-domain indicators encompass features related to the time between feature points, height differences among these points, and the waveform's area under the curve. The PPG waveform's second derivative, known as acceleration PPG, characterizes the waveform's contour curvature (Takazawa et al., 1998). The frequency-domain features are the power spectral densities (PSD) generated by the Fast Fourier Transform (FFT) and several derived values from the PSDs (Figure 3). Although studies typically use the heights of the peaks to represent the relative strengths of harmonics, due to the susceptibility of spike heights to random noise, this study uses the curve's area under the curve (approximating the interval's average value) for a more accurate representation of relative strength. For each sample, we use the median of the values obtained from all the BBIs to represent the sample.

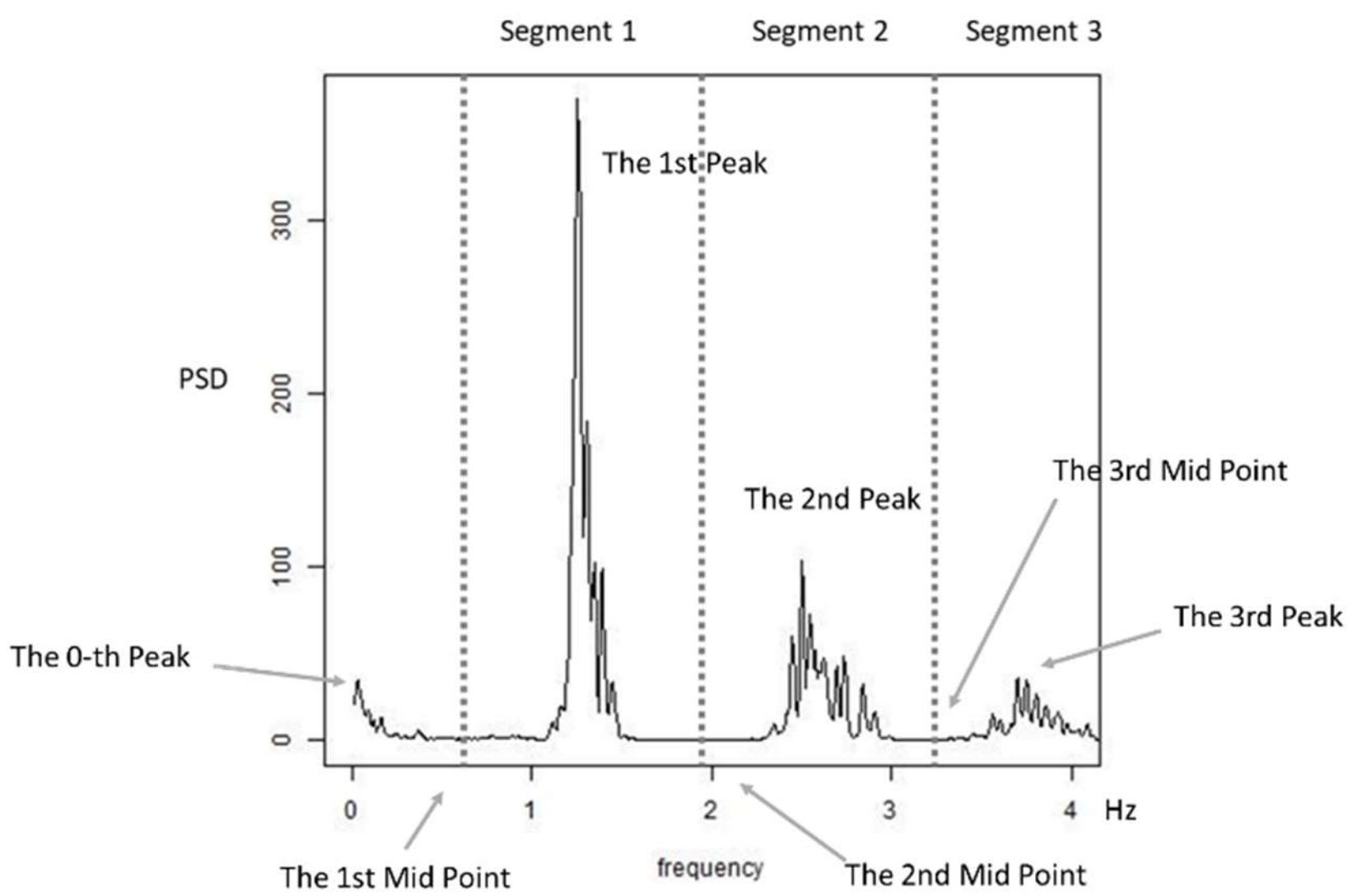

Figure 3. An illustration of the PSD generated by FFT and the segmentation.

TABLE 1 Waveform Features

| Name | Definition and calculation details |
| --- | --- |
| **Time and altitude-domain indicators** | |
| Peak-to-peak time (PPT) | The interval between the early systolic peak [ESP, also known as the left peak (LP) or the first peak (FP)] and the diastolic peak (DP). The right valley (RV) is the lowest point in the BBI, and the preceding BBI's RV is termed the left valley (LV). The diastolic peak [DP, or second peak (SP)], is the first point between ESP and RV with a zero first derivative and a negative second derivative. If no second peak exists between ESP and RV, indicating a consistently negative slope from LP to RV, the DP is identified as the point with the minimum second derivative between ESP and RV. |
| Reflection index (RI) | The ratio of the diastolic peak height (DPH) to the early systolic peak height (ESPH) |
| Stiffness index (SI) | The ratio of body height to PPT (Binder et al., 2008) (Du et al., 2015) |
| Crest time (CT) | The interval between the left valley (LV) and ESP (Korpas et al., 2009). |
| Notch time (NT) | The interval between LV to the dicrotic notch (DN). The lowest point between ESP and DP is the dicrotic notch (DN). In cases without a second peak, IP is designated as DN. |
| Diastolic time (DT) | The interval between ESP to the right valley (RV) (Teng & Zhang, 2003) |
| RCA | The ratio of CT to NT. |
| RDA | The ratio of NT to the combined duration of CT and DT. |
| Inflection point area (IPA) | The diastole to systole area ratio is calculated from the area under the curve from IP to RV (A2) relative to the area from LV to IP (A1) (Wang et al., 2009). The inflection point (IP), defined as the last point before DP where the first derivative shifts from positive to negative, is crucial in calculating the IPA. Points E or DN demarcate the diastolic and systolic areas (Charlton et al., 2018; Sharman et al., 2005), with A1 approximated as a polygon formed by LV, ESP, DN, and DNB, and A2 as a triangle comprising IP, DN, and RV, simplifying the calculation process. |
| **Acceleration PPG** | |
| A | The first local maximum of the second derivative, determined by the point with the largest second derivative before ESP. |
| B | The first local minimum of the second derivative, determined by the point with the smallest second derivative before ESP. |
| C | The second local maximum of the second derivative, determined by the point with the largest second derivative point B and point E. |
| D | The second local minimum of the second derivative, determined by the point with the smallest second derivative between point B and point E. |
| E | The third local maximum of the second derivative, determined by the point with the largest second derivative between ESP and point F. |
| F | The third local minimum of the second derivative, determined by the |
| | point with the smallest second derivative between the ESP and RV. |
| G | The fourth local maximum of the second derivative, determined by the point with the largest second derivative between point F and point H. |
| H | The fourth local minimum of the second derivative, determined by the point the smallest second derivative between point F and RV. |
| BA, EA, FA, GA | \|B/A\|, \|E/A\|, \|F/A\|, and \|H/A\|. Given that B/A, E/A, F/A, and H/A are negative, we converted them to their absolute values for a more intuitive interpretation. |
| Aging Index (AI) | BA– EA. AI is usually defined as BA – (CA+DA+EA) (Bortolotto et al., 2000). However, in this study, AI is defined as the difference between BA and EA only, due to difficulties in identifying points C and D with poor signal quality and the fact that the values of C and D are close to zero. |
| **Frequency-domain features** | |
| PSDi | The i-th relative power spectrum density. The continuous PSD curve is segmented using the midpoints between each peak. We define the strength of the i-th harmonic ($PSD_i$, starting from 1) as the area between the i-th midpoint and the (i+1)-th midpoint. Since the absolute values of each raw $PSD_i$ is influenced by the quality of the signal, the values are normalized as the relative values: $PSDi = \frac{raw\ PSD_i}{\sum_{j=1}^{6} raw\ PSD.j}$ |
| NHA | $1 - PSD1$ (Brown, 1999). |
| IHAR | 1-NHA/IPA (Brown, 1999) |

Initial analysis revealed that our waveforms were smoother than the theoretical model depicted by the green line in Figure 4, with a reflection index (RI) lower than values previously reported, around 0.85 (Panula et al., 2019). We primarily attribute this discrepancy to the auto-exposure adjustment feature of smartphone cameras. Intended to optimize image quality for human vision, this feature modifies pixel RGB values to maintain a visually comfortable average. However, since RGB values are limited to a 0 to 256 range, this adjustment may compress boundary values. Consequently, this compression results in a dampened waveform, evident from the average pixel values, particularly when pixels approach the RGB scale's extremities (Khanoka et al., 2004).

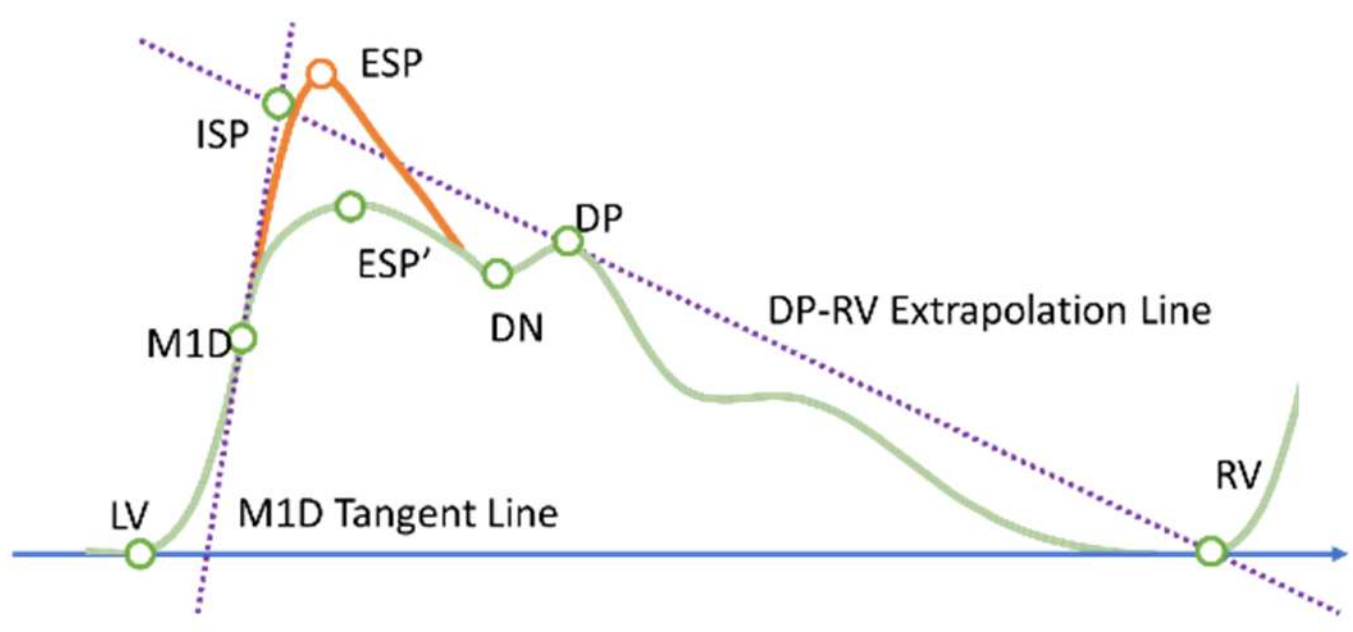

Figure 4 Theoretical pulse waveform (Fujime et al., 2012) vs. observed results gathered by

HH.

To address this issue, we introduce a new method for estimating the ESP height. Our method calculates the ESP height using the intersection point (ISP, shown in Figure 4) of the tangent line at the point of maximum first derivative and the extrapolation line connecting DP and RV. Subsequently, we define the Expected Reflection Index (ERI) as the ratio of the DP height to the ISP height. Furthermore, we introduce the Adaptive Reflection Index (ARI), set equal to ERI when the BA is less than or equal to 1, and equal to the RI when BA exceeds 1.

### 2.4.ECG Reference Comparison and Bland-Altman Analysis

Bland-Altman analysis is a method used for assessing the agreement between two techniques that measure the same target (Giavarina, 2015). It is recommended by the Artery Society for comparing non-invasive hemodynamic measurement devices (Wilkinson et al., 2010). We also adopted the standard set by The British Hypertension Association Standard (BHS), Association for the Advancement of Medical Instrumentation (AAMI), International Organization for Standardization (ISO), the European Society of Hypertension (ESH) and the American National Standards Institute (ANSI) in this study.

BHS requires that the BP estimation and measurement equipment should be graded according to the cumulative frequency error percentage of the estimated result value, which is Grade A (60% < 5 mmHg; 85% < 10 mmHg; 95% < 15 mmHg), Grade B (50% < 5 mmHg; 75% < 10 mmHg; 90% < 15 mmHg), grade C (40% < 5 mmHg; 65% < 10 mmHg; 85% < 15 mmHg)(O’Brien et al., 1993). The consensus of the AAMI, ISO, ESH and ANSI is that an error of ≤10 mm Hg (using an individual’s average of 3 BP readings versus a reference BP measurement method) and an estimated probability of that error of at least 85% is acceptable (Stergiou et al., 2018).

## 3.Results

### 3.1.Data Processing and Normalization

This study collected 766 waveform samples, approximately 6.03 per participant, and 229,165 Beat-to-Beat Intervals (BBIs), about 299.17 per sample, from 127 participants. Of these samples, 527 (69.8%) from 113 participants satisfied the validity criteria: a Smartphone PPG Signal Quality Index above 0.8(Liu et al., 2020), a minimum of 15 data points per heartbeat, and at least 100 seconds of collection time. Accounting for the established link between waveform features and heart rate (HR), all features were normalized to a standard HR of 75 beats per minute, following precedent in previous studies (Kelly et al., 2001). While considering the relationship between waveform features, height, and gender, we opted to include only height in our analysis, as advised by O’Rourke et al., (2001), standardizing features to a height of 170 cm, except for the Systolic Index (SI), which does not require height normalization. Initial analysis revealed positive skewness in acceleration PPG features and the frequency-domain feature index, prompting a log transformation for subsequent analysis. Additionally, outliers were removed by excluding data points exceeding 1.5 times the interquartile range from the first and third quartiles.

### 3.2.Statistical Analysis

We calculated the correlation coefficient for single variables showing significant linear associations with SBP, DBP, and SBP-DBP. Considering the study involved 28 variables, we applied the Bonferroni correction to reduce selection bias, setting the significance level for the p-value at $1.8 \times 10^{-3} (= 0.05/28)$. Our data showed that more than half of the normalized waveform indices were significantly correlated with at least one of the measures of BP (TABLE 2). RI, ERI, ARI, NT, IPA, EA, PSD2, and PSD6 were correlated with both SBP and DBP; PSD4 was correlated with SBP; PPT, PSD1, NHA, and IHAR were correlated with DBP; PSD4 and PSD6 were correlated with SBP-DBP. Body height (Height) was correlated with SBP, DBP, and SBP-DBP, while HR was correlated with SBP-DBP.

TABLE 2 Correlation coefficient, full regression model, stepwise regression model analysis results

| | **SBP** | | | **DBP** | | | **SBP-DBP** | | |
|---|---|---|---|---|---|---|---|---|---|
| Features | $r$ | Full | Stepwise | $r$ | Full | Stepwise | $r$ | Full | Stepwise |
| Height | $^{***}$0.45 | | | $^{***}$0.39 | | | $^{***}$0.35 | | |
| HR | -0.03 | | | 0.13 | | | $^{***}$-0.25 | | |
| Age | -0.08 | | | -0.01 | | | -0.14 | | |
| SI | 0.06 | $^{**}$-0.76 | $^{***}$-0.62 | 0.11 | -0.10 | | -0.10 | $^{***}$-1.51 | $^{***}$-0.84 |
| RI | $^{*}$0.15 | 0.06 | | $^{***}$0.22 | 0.15 | | -0.06 | -0.23 | $^{***}$-0.32 |
| ERI | $^{***}$0.24 | -0.20 | | $^{***}$0.30 | -0.02 | 0.11 | 0.00 | -0.32 | |
| ARI | $^{***}$0.26 | 0.16 | 0.16 | $^{***}$0.33 | $^{*}$0.24 | $^{***}$0.29 | 0.01 | -0.01 | |
| PPT | -0.13 | $^{**}$-1.21 | $^{***}$-0.73 | $^{*}$-0.16 | -0.35 | | 0.03 | $^{***}$-1.87 | $^{***}$-0.73 |
| AI | 0.13 | $^{**}$-1.43 | $^{**}$-0.97 | 0.09 | -1.06 | | 0.13 | $^{**}$-1.56 | |
| CT | -0.08 | 1.79 | $^{*}$1.07 | -0.04 | -0.09 | $^{**}$0.21 | -0.10 | $^{**}$4.79 | |
| NT | $^{***}$-0.21 | -0.43 | | $^{***}$-0.21 | 0.31 | | -0.08 | $^{*}$-1.89 | |
| DT | 0.05 | -0.27 | $^{***}$-0.46 | 0.01 | -0.17 | | 0.13 | -0.22 | $^{**}$-0.23 |
| IPA | $^{***}$0.19 | 0.88 | $^{***}$0.94 | $^{***}$0.22 | 0.57 | | 0.02 | 0.95 | $^{***}$0.89 |
| RCA | 0.05 | -2.46 | $^{*}$-2.37 | 0.08 | -0.59 | | -0.05 | $^{**}$-4.59 | |
| RDA | -0.09 | -1.13 | $^{**}$-1.47 | -0.13 | -1.17 | | 0.03 | -0.13 | |
| BA | 0.04 | $^{**}$1.23 | $^{**}$1.06 | 0.03 | $^{*}$1.00 | | 0.07 | $^{**}$1.16 | |
| EA | $^{**}$-0.16 | $^{*}$-0.46 | -0.28 | $^{**}$-0.16 | -0.34 | | -0.05 | $^{*}$-0.48 | |
| FA | -0.12 | $^{*}$-0.31 | $^{*}$-0.23 | -0.06 | $^{*}$-0.31 | $^{**}$-0.21 | -0.11 | -0.11 | |
| GA | 0.00 | $^{**}$0.28 | 0.12 | 0.07 | $^{*}$0.25 | $^{*}$0.16 | -0.09 | 0.15 | |
| HA | 0.09 | -0.09 | | 0.12 | -0.06 | | -0.01 | -0.10 | |
| PSD1 | 0.11 | $^{*}$1.24 | | $^{*}$0.16 | -0.04 | | -0.05 | $^{***}$2.50 | $^{***}$0.85 |
| PSD2 | $^{***}$-0.19 | 0.27 | | $^{***}$-0.25 | -0.26 | | 0.03 | $^{*}$0.79 | |
| PSD3 | -0.03 | $^{*}$0.57 | | -0.05 | -0.00 | | 0.04 | $^{***}$1.05 | $^{***}$0.38 |
| PSD4 | $^{*}$0.14 | 0.08 | | 0.09 | -0.13 | | $^{**}$0.17 | $^{*}$0.31 | |
| PSD5 | 0.09 | 0.05 | | 0.07 | -0.02 | | 0.11 | 0.13 | |
| PSD6 | $^{***}$0.20 | $^{***}$0.38 | $^{***}$0.32 | $^{**}$0.17 | $^{**}$0.33 | $^{***}$0.29 | $^{**}$0.18 | $^{**}$0.32 | $^{***}$0.37 |
| NHA | -0.10 | $^{*}$0.51 | | $^{**}$-0.18 | 0.18 | | 0.11 | $^{**}$0.81 | $^{***}$0.59 |
| IHAR | -0.12 | $^{**}$0.95 | $^{***}$0.95 | $^{***}$-0.19 | $^{**}$0.86 | | 0.07 | $^{*}$0.63 | $^{***}$0.68 |
| Adj-$r^2$ | | $^{***}$0.27 | $^{***}$0.27 | | $^{***}$0.21 | $^{***}$0.18 | | $^{***}$0.30 | $^{***}$0.27 |
| n | | 358 | 358 | | 358 | 358 | | 350 | 350 |

Note: *: $p<1.8\times10^{-3}$; **: $p<4\times10^{-4}$; ***: $p<4\times10^{-5}$ for correlaton coefficient analysis; *: p<0.05; **: p<0.01; ***: p<0.001 for full and stepwise regression model; $\boldsymbol{r}$ the correlation coefficient; Full: the $\beta$ value of the full regression model; Stepwise: the $\beta$ value of the stepwise regression model

The full-model multiple linear regression (MLR) and stepwise analysis elucidated the multivariate relationships between waveform characteristics and blood pressure, demonstrating statistical significance (P<0.05). The full-model yielded adjusted R-square values of 0.27 for SBP, 0.21 for DBP, and 0.30 for SBP-DBP. In predicting SBP, indices such as SI, PPT, AI, BA, EA, FA, GA, PSD1, PSD3, PSD6, NHA, and IHAR emerged as significant predictors (P<0.05). For DBP prediction, ARI, BA, FA, GA, PSD6, and IHAR were identified as significant (p<0.05). Additionally, SI, PPT, AI, CT, NT, RCA, BA, EA, PSD1, PSD2, PSD3, PSD4, PSD6, NHA, and IHAR showed significance in predicting SBP-DBP (p<0.05).

Employing stepwise regression models and the Akaike Information Criterion (AIC) for feature selection (Venables & Ripley, 2013), key predictors were identified using the stepAIC package in R, utilizing both forward and backward selection methods. The analysis resulted in adjusted $r^2$ values of 0.27 for SBP, 0.18 for DBP, and 0.27 for SBP-DBP. PSD6 stood out as a crucial predictor for SBP, DBP, and their difference. Additionally, CT and FA were significant for SBP and DBP, whereas SI, PPT, DT, IPA, and IHAR were pertinent for SBP and SBP-DBP.

Beyond mere statistical tests, this study utilized 10-fold cross-validation to evaluate prediction accuracy. Using waveform indices alone, MLR achieved an $r^2$ of 0.25 for SBP, 0.15 for DBP, and 0.26 for SBP-DBP (Figure 5). The inclusion of non-waveform variables, such as body height and heart rate, significantly improved prediction accuracy, resulting in $r^2$ of 0.40 for SBP, 0.33 for DBP, and 0.32 for the SBP-DBP difference. Furthermore, the study explored the influence of outliers on accuracy, uncovering a significant reduction in all MLR correlation coefficients attributable to outliers, highlighting their effect on predictive performance.

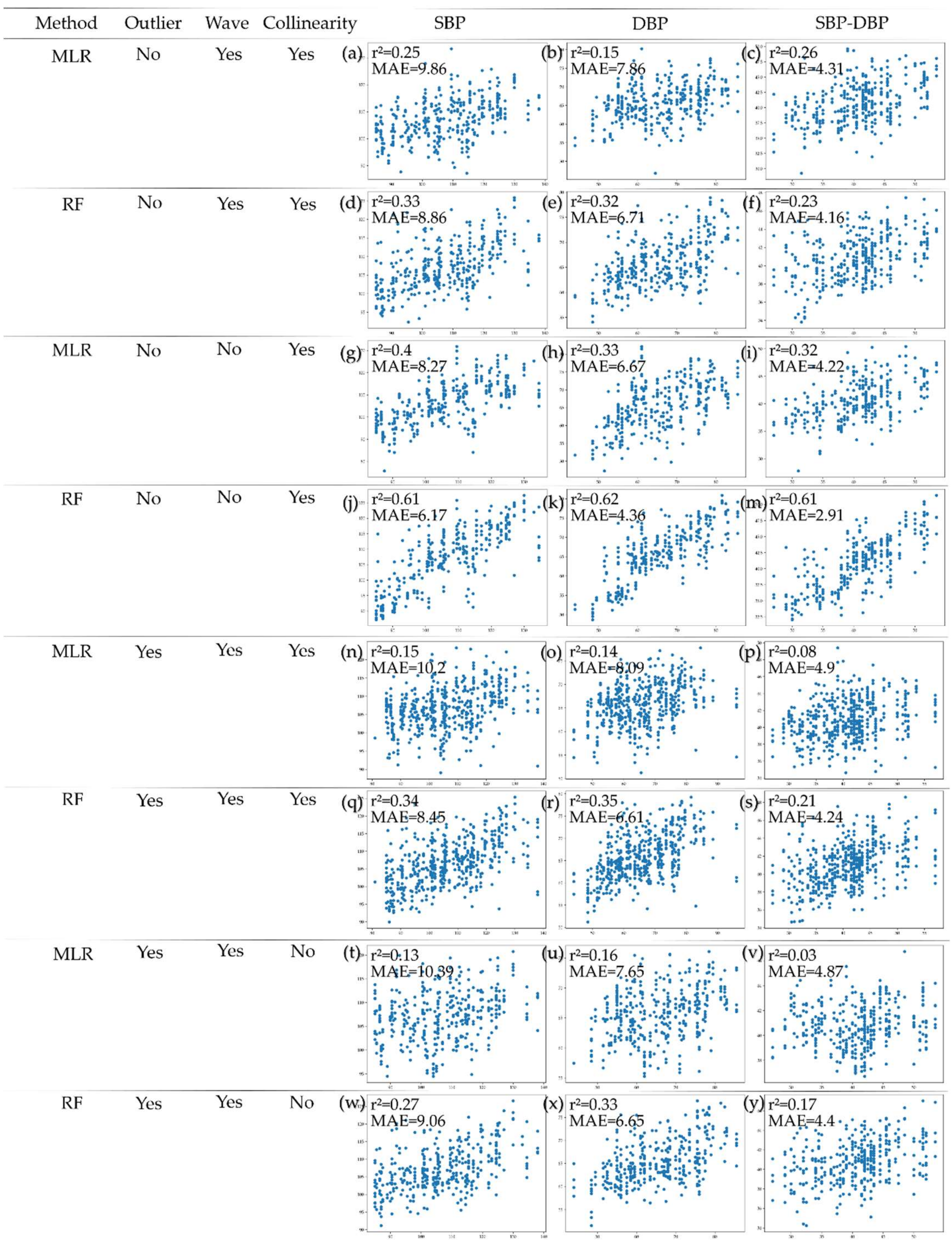

Figure 5. The correlation coefficients of the predicted BP (y-axis) and the referencing BP (x-axis). Wave: using only waveform features; Outlier: using data that contain outliers (n=513); Collinearity: using features with high collinearity. For (n),(o),(p),)(q),(r) and (s), the vertical axis was trimmed so some of the outliers of prediction are not shown on the graphs.

High correlations among predictors may result in counterintuitive, uninterpretable, misleading, and unstable outcomes in statistical analyses (Chan et al., 2022). Our research identified significant collinearity among waveform features. To address this issue, we applied Dormann et al. (2013)'s approach by removing the variable with the lower correlation from each pair exhibiting a correlation coefficient above 0.7, leaving us with a refined set of features: ARI, PPT, NT, BA, EA, FA, GA, HA, PSD2, PSD3, PSD4, and PSD6 (TABLE 3). Despite a correlation of only 0.6 between RI and ARI, their conceptual similarity prompted the exclusion of RI in favor of ARI. Following collinearity adjustments, the adjusted $r^2$ values for our MLR analysis decreased to 0.15 for SBP, 0.18 for DBP, and 0.05 for the SBP-DBP difference, as detailed in TABLE 3. The stepwise regression analyses also showed a comparable decrease in adjusted $r^2$ values.

TABLE 3 Linear Regression Analysis and Stepwise Regression Analysis with Collinearity Removed

| | SBP $\beta$ | | DBP $\beta$ | | SBP-DBP $\beta$ | |
|---|---|---|---|---|---|---|
| Features | Full | Stepwise | Full | Stepwise | Full | Stepwise |
| ARI | $^{**}$0.20 | $^{***}$0.19 | $^{***}$0.29 | $^{***}$0.23 | -0.03 | |
| PPT | $^{**}$-0.31 | $^{***}$-0.20 | $^{**}$-0.33 | $^{***}$-0.28 | -0.08 | |
| NT | -0.07 | | 0.00 | | -0.14 | $^{*}$-0.13 |
| BA | 0.06 | | 0.05 | 0.10 | 0.06 | |
| EA | 0.05 | | 0.03 | | 0.07 | |
| FA | -0.18 | $^{**}$-0.19 | -0.15 | | -0.05 | $^{**}$-0.15 |
| GA | 0.06 | 0.09 | 0.17 | | -0.19 | |
| HA | 0.03 | | -0.05 | | 0.14 | |
| PSD2 | 0.08 | | 0.03 | | 0.13 | 0.10 |
| PSD3 | -0.12 | | -0.08 | $^{*}$-0.12 | -0.18 | |
| PSD4 | 0.05 | | 0.00 | | 0.11 | |
| PSD6 | $^{**}$0.22 | $^{***}$0.24 | $^{**}$0.22 | $^{***}$0.23 | 0.15 | $^{**}$0.15 |
| Adjusted $R^2$ | $^{***}$0.15 | $^{***}$0.15 | $^{***}$0.18 | $^{***}$0.18 | $^{**}$0.05 | $^{***}$0.05 |
| $n$ | 384 | 384 | 384 | 384 | 375 | 375 |

Note: *: *P*<.05; **: *P*<.01, ***: *P*<.001.

### 3.3. Machine Learning Analysis

To improve prediction accuracy, we utilized the Random Forest (RF) model, validated through 10-fold cross-validation. The RF model outperformed the MLR in predicting SBP ($r^2 = 0.33$, Figure 5), DBP ($r^2 = 0.32$), and SBP-DBP ($r^2 = 0.23$). Incorporating non-waveform factors like height and heart rate significantly improved accuracy for SBP ($r^2 = 0.61$), DBP ($r^2 = 0.62$), and SBP-DBP ($r^2 = 0.61$). However, after collinearity adjustments with a reduced set of variables, accuracy modestly decreased for SBP ($r^2 = 0.27$), DBP ($r^2 = 0.33$), and SBP-DBP ($r^2 = 0.17$).

We also conducted SHAP analysis using the 'shap' package in Python. For MLR, the five most influential waveform features for predicted values were RCA, CT, BA, AI, and PSD1 for SBP (Figure 6); RDA, NT, CT, BA and IHAR for DBP; CT, RCA, PSD1, NT, and PPT for SBP-DBP. In contrast, when employing the Random Forest (RF) model to predict SBP, DBP, and SBP-DBP, the top five feature sets identified were PPT, PSD6, EA, IHAR, and RDA for SBP; PPT, EA, ARI, PSD6, and PSD4 for DBP; NHA, PPT, BA, PSD4, and NT for SBP-DBP. Notably, compared to MLR, the RF model tends to focus on a narrower set of features, with some being predictive for both SBP and DBP, highlighting a significant difference in feature importance between the MLR and RF models.

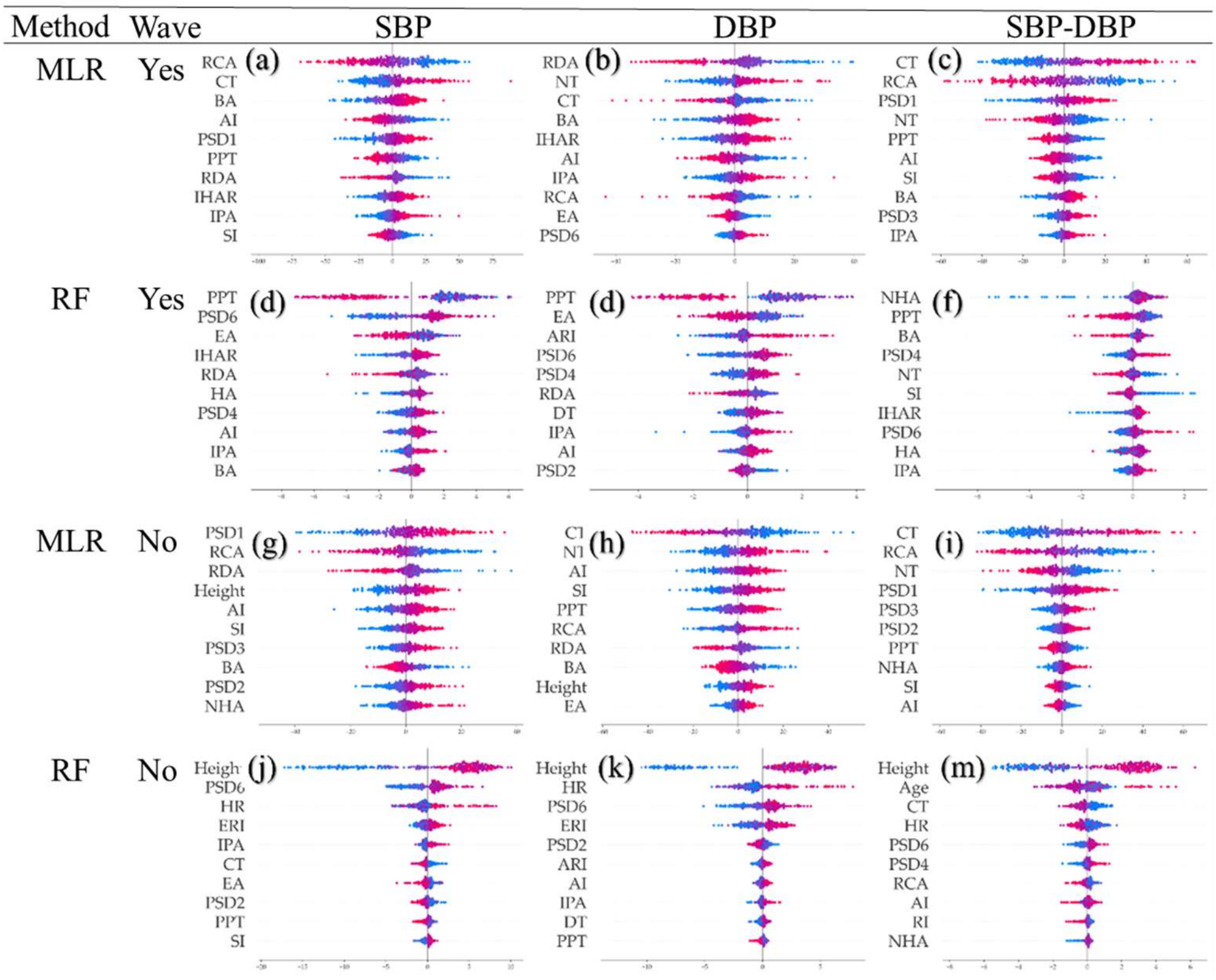

Figure 6. The bee-swarm plot illustrates the SHAP values for MLR and RF predictions, with each dot representing the SHAP value of an individual sample. Variables are displayed in descending order of their global feature importance. The 'Wave' category differentiates between models: 'Yes' signifies models that only include waveform features, while 'No' indicates the inclusion of non-waveform features. In this analysis, all outliers have been excluded, yet collinearity among variables has not been addressed.

Integrating non-waveform variables like height, age, and HR into the MLR analysis indicated that height significantly predicted both SBP and DBP, whereas age and HR did not prove to be significant predictors in this model. In stark contrast, employing the RF model, all non-waveform variables, encompassing height, age, and HR, were deemed crucial for predicting SBP, DBP, and the SBP-DBP difference.

### 3.4. ECG Reference Comparison and Bland-Altman Analysis

The Bland-Altman analysis indicated that most of differences between predicted and reference results fell within the defined upper and lower limits. Nevertheless, regression analyses between predicted and reference results demonstrated significance ($p < 0.05$, Figure 7). Consequently, Bland-Altman plots revealed a systematic bias in both MLR and RF models, wherein the predicted values consistently underestimated the true values, especially as the actual values increased.

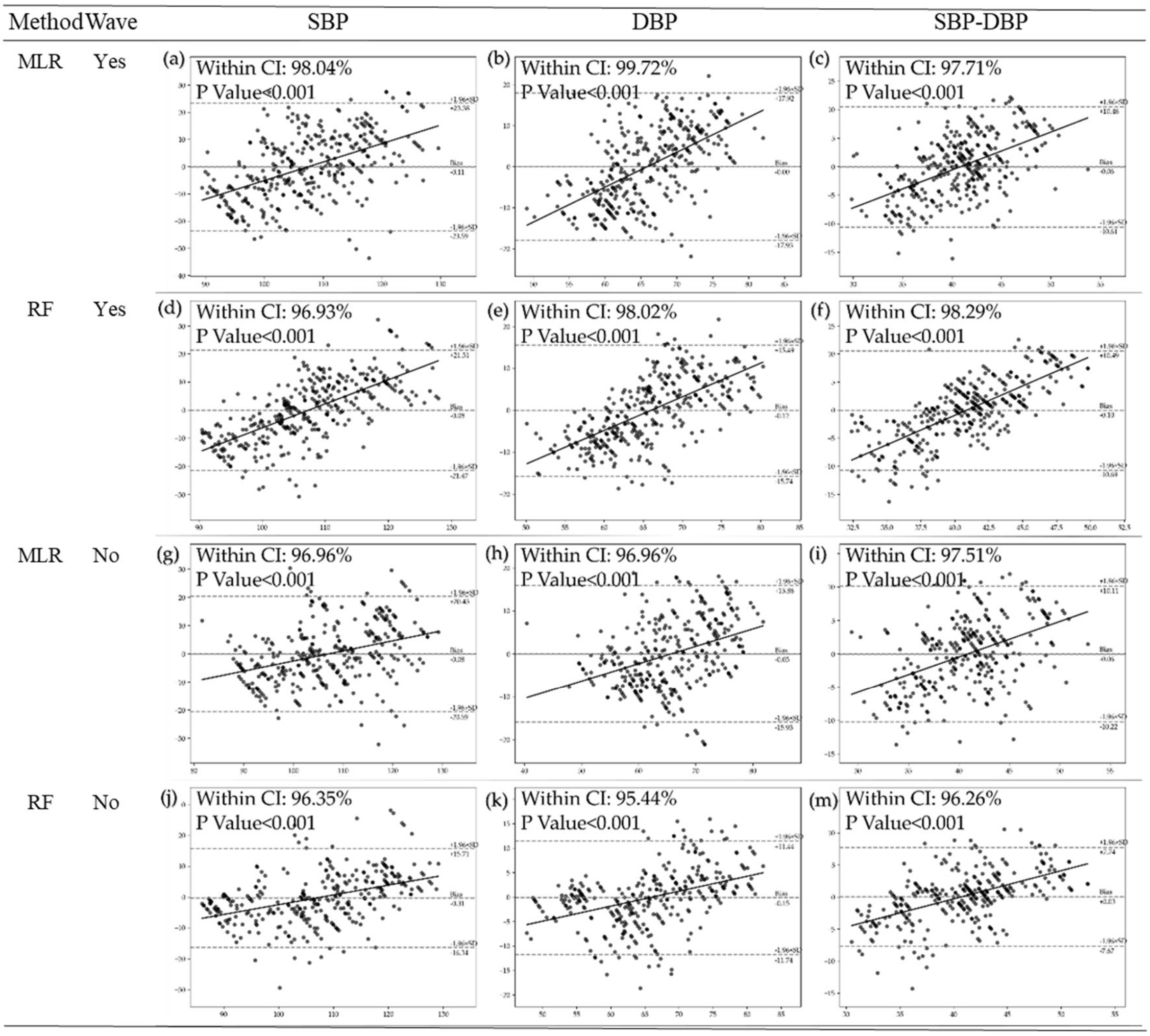

Figure 7. The Bland–Altman plot for the predicted results and the actual results of measuring BP. For each plot, the x axis indicates the average of the two results, and the y axis indicates the difference between the two results. Wave: Yes: only include wave features; No: include non-waveform features. All outliers were removed and collinearity is not removed.

In comparing predictive outcomes with reference data, our analysis focused on collinear waveform characteristics extracted from datasets devoid of outliers. This analysis indicated that the distribution of prediction absolute errors (AE) primarily fell within a 15 mmHg range for SBP, DBP, and the SBP-DBP difference. Despite this relatively limited error range, neither the MLR nor the RF models achieved the acceptable standards set by AAMI, ISO, ESH, and ANSI. Specifically, the RF model's performance for DBP was classified as grade C, and for SBP-DBP as grade B, according to BHS criteria. In contrast, the MLR model's performance for SBP-DBP was rated as grade C by the same standards.

Table 4. Bland-Altman Analysis

| The difference of the two results | SBP | DBP | SBP-DBP |
|---|---|---|---|
| MLR | | | |
| Mean of the error (mmHg) | -0.11 | -0.03 | -0.06 |
| SD of the error (mmHg) | 11.98 | 9.53 | 5.38 |
| MAE (mmHg) | 9.86 | 8.01 | 4.31 |
| SD of the AE (mmHg) | 6.78 | 5.15 | 3.21 |
| n. of data with AE within 0-5 mmHg | 97(27.09%) | 116(32.40%) | 222(63.43%) |
| n. of data with AE within 6-10 mmHg | 104(29.05%) | 116(32.40%) | 106(30.29%) |
| n. of data with AE within 11-15 mmHg | 75(20.95%) | 96(26.82%) | 20(5.71%) |
| n. of data with AE > 15 mmHg | 82(22.91%) | 30(8.38%) | 2(0.57%) |
| RF | | | |
| Mean of the error (mmHg) | -0.23 | 0.02 | -0.08 |
| SD of the error (mmHg) | 10.92 | 8.09 | 5.47 |
| MAE (mmHg) | 8.91 | 6.68 | 4.26 |
| SD of the AE (mmHg) | 6.30 | 4.54 | 3.43 |
| n. of data with AE within 0-5 mmHg | 118(32.96%) | 155(43.30%) | 232(66.29%) |
| n . of data with AE within 6-10 mmHg | 97(27.09%) | 127(35.47%) | 89(25.43%) |
| n. of data with AE within 11-15 mmHg | 88(24.58%) | 57(15.92%) | 29(8.29%) |
| n. of data with AE >15 mmHg | 55(15.36%) | 19(5.31%) | 0(0.00%) |
| t-test for the mean differences of AE | P=0.02 | P<0.01 | P=0.81 |
| Cohen's d | 0.145 | 0.274 | 0.015 |

## 4.Discussion

### 4.1.1Principal Findings

#### 4.1.1.The accuracy-interpretability dilemma

Our data supported the proposed strategy that enhancing data preprocessing can significantly improve the prediction accuracy of both more interpretable traditional statistical methods and less interpretable machine learning methods. The SHAP analysis further improves the interpretability of machine learning results and therefore mitigates the accuracy-interpretability dilemma.

The MAE for traditional statistical methods, specifically MLR, was reduced to 9.86±6.78 for SBP, 8.01±5.15 for DBP, and 4.31±3.21 for the SBP-DBP difference, respectively after we eliminated corrupted data and outliers, precisely normalized measurements, and accounted for the limitations inherent in the automatic exposure of smartphone cameras. Conversely, the machine learning models registered an MAE of 8.91±6.30 for SBP, 6.68±4.54 for DBP, and 4.26±3.43 for the SBP-DBP difference, respectively. The $r^2$ (excluding outliers, non-waveform features, or collinearity), improved from 0.25, 0.15, and 0.26 for SBP, DBP, and the SBP-DBP difference utilizing MLR, to 0.33, 0.32, and 0.23 for those respective metrics using RF. The results demonstrate the superiority of machine learning techniques while also highlighting the usefulness of traditional MLR methods after careful signal preprocessing.

Although our results validate the proposed strategy, the error values in our study are noticeably higher non SPW-BP studies. For example, Gao et al. (2016) achieved an MAE for DBP and SBP of 4.6 ± 4.3 mmHg and 5.1 ± 4.3 mmHg, respectively, using discrete wavelet transform for feature extraction from oximeter PPG signals and nonlinear SVM for prediction. Similarly, Dey et al. (2018) reported MAEs for DBP and SBP of 5.0 ± 6.1 mmHg and 6.9 ± 9.0 mmHg, respectively, after analyzing 233 time-frequency domain features with a Infrared heart rate sensor on a Samsung Galaxy S6 and identifying Lasso regression as the most accurate. Therefore, there is still much room for improvement in smartphone camera-based methods for acquiring better signal quality.

### 4.1.2.Feature Importance

There is no consensus yet on which waveform features most significantly related to BP measurements. Some research posits that waveform characteristics are predominantly associated with DBP (Brown, 1999), whereas other studies argue for a stronger correlation with SBP (Cameron et al., 1998). Additionally, several studies suggest a relationship between waveform features and both SBP and DBP (Kelly et al., 2001). Our correlation analysis revealed a significant association between various BP measurements and multiple investigated waveform features.

Time and altitude domain indices were found to be related to both SBP and DBP in previous studies. Arterial aging was usually positively correlated with RI and inversely correlated with PPT (Millasseau et al., 2006). Several studies have found a significant positive correlation between SI and pulse wave velocity (Padilla et al., 2006). CT and DT have been observed to be negatively correlated with BP measures (Teng & Zhang, 2003). Consistent with prior studies (Padilla et al., 2006), the RI showed positive relationships with SBP and DBP in our data. However, the link between SI, CT and DT, and the BP measures were not significant in our data.

The proposed remedy for the dampened waveform yielded promising results. ARI exhibited a notably stronger correlation with both SBP and DBP compared to RI. Furthermore, during the feature selection process, ARI was more frequently selected by MLR, stepwise regression, and RF as part of the optimal feature sets for BP prediction. This outcome suggests that adjusting for issues created by the autoexposure function of smartphone cameras can significantly enhance the predictability of BP measurements.

Research indicates that acceleration PPG indices are associated with arterial stiffness, age, risk of heart attack, and the distensibility of the peripheral artery (Elgendi, 2012). Compared to individuals with normal BP, those with hypertension typically exhibit significantly lower BA and CA ratios, but higher DA and EA ratios (Simek et al., 2005). In our dataset, acceleration PPG features were expectedly good indicators of BP. However, the linear correlation coefficients of BP measures to BA, FA, GA, and HA were not significant, while EA was negatively correlated with BP measures. Using full-model MLR and stepwise regression, BA and GA were found to be positively correlated with BP, whereas EA and FA were negatively correlated with both SBP and DBP. The inconsistency with previous studies might be attributed to the dampened waveform issue discussed earlier. Future research should further explore this phenomenon.

Although early cardiovascular studies have investigated the frequency-domain features of waveforms (Christensen & Børgesen, 1989), this study is, as far as we know, the first study that links waveform features to BP. The results of our study suggest that PSD2, PSD4, and PSD6 may be associated with BP. In SHAP analysis, PSD6 was essential for RF, while PSD1 was significant for MLR. Since frequency-domain features are highly correlated to many time-domain features, these results were expected.

In our analysis, we explored the difference in prediction accuracy between models using only waveform features versus those incorporating non-waveform features such as gender, heart rate, and body height. As expected, including these non-waveform features significantly improved prediction accuracy. This approach was motivated by prior studies that often indiscriminately add non-waveform features to machine learning models, resulting in high accuracy for pulse waveform-based BP prediction (Frey et al., 2022). However, this high accuracy could be due to the inclusion of unrelated variables, underscoring the importance of interpretability in machine learning to ensure conclusions are valid and to avoid misleading results.

Consensus on the best feature selection method remains elusive. In multivariate analyses, stepwise regression and MLR often identify different feature sets. While some researchers prefer the Akaike Information Criterion (AIC) (Halsey, 2019), others regard *p*-values, confidence intervals, and information-theoretic criteria as various ways to summarize statistical information (Murtaugh, 2014). Our study found that univariate and multivariate analyses yielded different significant feature sets, with some features important for BP showing low importance in SHAP analysis. This underscores the complexity of feature selection in machine learning and statistical modeling. While the goal of our study was not to determine the best feature set, these results opened avenues for future research.

## 4.2.Limitations

### 4.2.1.Low Signal Quality

Our study highlighted the potential of smartphones for BP prediction, though it did not meet the Bland-Altman analysis threshold. The insufficient agreement may stem from low signal quality and the inherent randomness of PPG signals, affected by factors like skin tone, age, gender, and environmental conditions (Fine et al., 2021). External influences and variables such as ambient light and motion artifacts due to unstable holding of the phone complicate signal consistency. Additionally, the typical smartphone camera frame rate of 30 frames per second (Liu et al., 2020) may be inadequate for accurate BP prediction, considering heart rate variability studies recommend a minimum of 125 Hz (Laborde et al., 2017). Despite this, our findings suggest that a 30 frames per second rate might suffice for valid BP estimation, indicating a need for further research to establish a standard frame rate for SPW-BP predictions.

Researchers have already begun to address the aforementioned issues through various methods. In terms of improving signal acquisition quality, studies have proposed detecting better color channels and regions of interest (ROI) (Nam & Nam, 2017), detecting noise in samples (Bashar et al., 2018), filtering high-quality samples based on signal quality standards (Liu et al., 2020), and using smartphones with higher sampling rates. In data analysis and processing, recent LED-based PPG research has introduced various deep learning models, such as CNN and LSTM (Tanveer & Hasan, 2019), which have been effective in reducing the MAE to $4.06 \pm 4.04$ and $3.33 \pm 3.42$ for SBP and DBP, respectively (Eom et al., 2020).

### 4.2.2.Collinearity Reduction

Collinearity is a key challenge in data analysis, with various mitigation methods available

(Chan et al., 2022). Principal Component Analysis and ridge regression are often used, though they may reduce result interpretability. Selecting representative variables can introduce bias and affect accuracy (Dormann et al., 2013). In our study, we used a simple variable removal method to increase interpretability and minimize bias, successfully reducing collinearity but decreasing the adjusted $r^2$ of our models. This underscores the trade-off between model interpretability and accuracy, illustrating the complexity of addressing collinearity.

### 4.2.3.The True 2nd Peak

Several important time and altitude-domain indicators, such as PPT, RI, and SI, are based on the information of the second peak. However, there is no consensus in the literature on PPG-based waveform analysis regarding the definition of the second peak. And, we believe the discrepancy might jeopardize the whole academic studies of wave form-based cardiovascular analysis.

In aortic pressure wave studies, RI is typically defined as the ratio between the height of the reflected wave and the original wave (Xiao et al., 2018). The waveform generally consists of three distinct peaks: early systolic peak, late systolic peak, and diastolic peak (Kim et al., 2014). The first arrival of the reflected wave results in the late systolic peak, whereas the diastolic peak is caused by a second reflected wave (Baruch et al., 2011).

Some studies use the late systolic peak as the second peak (Melenovsky et al., 2007; Panula et al., 2019), and define the radial augmentation index or RI as the ratio of this peak to the early systolic peak. Others consider the diastolic peak the second peak (Djeldjli et al., 2021). Several studies also define the radial augmentation index as the ratio between the late systolic peak and early systolic peak and the RI as the ratio between the diastolic peak and early systolic peak (Peltokangas et al., 2014). Some studies do not base their choice of the second peak on the physiological knowledge of the waveform; instead, they identify the second peak as the most prominent peak on the waveform following the initial peak (Millasseau et al., 2002).

Apart from the conceptual inconsistency, there is no consensus on the technical method for identifying the location of the second peak. Chowienczyk and colleagues (1999) determined the position of the second peak using only the first derivative, Elgendi (2012) used the second derivative, and Melenovsky and colleagues (2007) proposed using the peaks of the fourth derivative. Alty and colleagues (2007) suggested using an inflection point as an alternative when no clear second peak can be identified on the waveform.

In this study, due to the limited frame rate of the smartphone camera and substandard signal quality, we found it challenging to effectively identify three peaks in the waveform. As a result, we followed previous studies and used the second waveform peak, which is prominently visible on the waveform, as the second peak (Millasseau et al., 2002). While definitions may vary, both the late systolic peak and the diastolic peak are induced by reflected waves. Given that the arrival time of reflected waves is influenced by the stiffness of blood vessels and that arteriosclerosis is strongly correlated with blood pressure, it's reasonable to use either the diastolic peak or the late systolic peak for estimating blood pressure. However, while the general directions may be similar, waveform features can differ significantly in magnitude. Future studies should exercise caution when making comparisons.

#### 4.2.4. Limited participant diversification

Assessing BP through waveforms fundamentally relies on how vascular sclerosis affects the speed of reflected waves, thereby influencing BP. However, the factors influencing BP extend beyond vascular sclerosis. For instance, aging introduces changes that impact the PPG signal, including increased collagen deposition and skin thinning. These changes, in turn, affect BP and light transmission through the skin (Frey et al., 2022). Therefore, SPW-BP may not be robust against age, and the generalizability of these findings warrants further analysis.

## 5.Conclusion

Smartphone-based BP measurement, valued for its convenience and cost-effectiveness, is poised to enable non-invasive, continuous BP monitoring. Additionally, camera-based contact HR measurement methods are pivotal in advancing contactless technologies, broadening the scope of health-monitoring capabilities (Liu, Liu, Zhong, Ma, et al., 2024). Actualizing this potential requires resolving the accuracy-interpretability dilemma, necessitating improvements in predictive accuracy while ensuring outcomes are clear and understandable.

This study provides evidence towards addressing the identified challenges, with contributions detailed as follows:

First, through enhanced signal preprocessing and feature engineering, including the elimination of corrupted data and outliers, appropriate standardization of measurements, and accommodation for smartphone camera exposure limitations, this study demonstrates a method to mitigate the quality issues inherent in Smartphone PPG data. Traditional statistical techniques have elucidated the relationship between waveform characteristics and blood pressure, thereby elevating the precision of waveform predictions and their medical validity.

Second, this study deepens our understanding of SPW-BP through a comprehensive analysis of the relationship between various cardiac waveform indicators and BP, notably introducing the frequency-domain features in PPG-based BP research for the first time. This method seeks not only to improve prediction accuracy but also to enrich our comprehension of blood pressure from physiological and psychological standpoints.

Third, this study introducing interpretable machine learning to SPW-BP research addresses the historical challenge of explainability in machine learning-based studies, marking a novel advancement in the field. Fourth, the incorporation of Bland-Altman Analysis and AAMI and BHS standards into SPW-BP research, for the first time, highlights existing disparities between current SPW-BP applications and accepted medical standards, thereby guiding future research directions. Although our results show that SPW-BP has not yet reached an acceptable standard in medicine, our analysis makes researchers aware of the current gap and provides direction for future efforts.

Finally, despite the nascent stage of SPW-BP research, characterized by a scarcity of evidence and data, this study's collection of 766 samples from 127 subjects substantiates the practicality of SPW-BP and suggests pathways for continued investigation.

## Acknowledgements

This work was supported by the Shenzhen R & D Sustainable Development Funding [KCXFZ20230731093600002], Guangdong Digital Mental Health and Intelligent Generation Laboratory [2023WSYS010], Guangdong Philosophy and Social Science Foundation [GD23SQXY01], Shenzhen Education Science Policy Research Project [ZDZC23013], the National Natural Science Foundation of China [32371121] and Guangdong Medical Research Foundation [B2024262].

## Conflicts of Interest

none declared.

## Abbreviations

ARI : adaptive reflection index

BI: beat-to-beat interval

BP: blood pressure

CT: crest time

DBP: diastolic blood pressure

DN: dicrotic notch

DP: diastolic peak

DT: Diastolic time

DPH: height of the DP

ERI: expected reflection index

ESP: early systolic peak

ESPH: height of the ESP

FP: first peak

FFT: fast Fourier transform

HH: Heartily Happy

IP: inflection point

IPA: Inflection point area

LP: left peak

LV: left valley

LSTM: Long Short Term Memory

MAE: mean absolute error

MLR: multiple linear regression

NT: Notch time

PPG: photoplethysmography

PPT: peak-to-peak time

PSD: power spectral densities

PWA: pulse waveform analysis

RF: random forest

RI: reflection index

SBP: systolic blood pressure

SHAP: SHapley Additive exPlanations

SI: stiffness index

SP: second peak

SPW-BP: smartphone PPG-based waveform analysis for blood pressure prediction